\RequirePackage{ifpdf}
\ifpdf % We are running pdfTeX in pdf mode
\documentclass[pdftex]{sigma}
\else
\documentclass{sigma}
\fi

\numberwithin{equation}{section}

\begin{document}

\renewcommand{\PaperNumber}{096}

\FirstPageHeading

\ShortArticleName{Lagrangian Approach to Dispersionless KdV
Hierarchy}

\ArticleName{Lagrangian Approach to Dispersionless\\
 KdV Hierarchy}

\Author{Amitava CHOUDHURI~$^{\dag^1}$, B.
TALUKDAR~$^{\dag^1\dag^2}$ and U. DAS~$^{\dag^3}$}

\AuthorNameForHeading{A.~Choudhuri, B.~Talukdar and U.~Das}

\Address{$^{\dag^1}$~Department of Physics, Visva-Bharati
University, Santiniketan 731235, India}
\EmailDD{\href{mailto:amitava_ch26@yahoo.com}{amitava\_ch26@yahoo.com}}

\Address{$^{\dag^2}$ Department of Physics, Visva-Bharati
University, Santiniketan 731235, India}
\EmailDD{\href{mailto:binoy123@bsnl.in}{binoy123@bsnl.in}}

\Address{$^{\dag^3}$~Abhedananda Mahavidyalaya, Sainthia 731234,
India}

\ArticleDates{Received June 05, 2007, in f\/inal form September
16, 2007; Published online September 30, 2007}

\Abstract{We derive a Lagrangian based approach to study the
compatible Hamiltonian structure of the dispersionless KdV and
supersymmetric KdV hierarchies and claim that our treatment of the
problem serves as a very useful supplement of the so-called
$r$-matrix method. We suggest specif\/ic ways to construct results
for conserved densities and Hamiltonian operators. The Lagrangian
formulation, via Noether's theorem, provides a method to make the
relation between symmetries and conserved quantities more precise.
We have exploited this fact to study the variational symmetries of
the dispersionless KdV equation.}

\Keywords{hierarchy of dispersionless KdV equations; Lagrangian
approach; bi-Hamiltonian structure; variational symmetry}

\Classification{35A15; 37K05; 37K10}

\section{Introduction}
The equation of Korteweg and de Vries or the so-called KdV
equation
\begin{gather*}
u_t=\tfrac{1}{4}u_{3x}+\tfrac{3}{2}uu_{x}%\label{eq1}
\end{gather*}
in the dispersionless limit \cite{zakharov}
\begin{gather*}
\frac{\partial}{\partial t}\rightarrow \epsilon  \frac{\partial}{\partial t}\qquad {\rm and}\qquad \frac{\partial}{\partial x}\rightarrow \epsilon  \frac{\partial}{\partial x}\qquad {\rm with}\quad \epsilon \rightarrow 0 %\label{eq2}
\end{gather*}
reduces to
\begin{gather}
u_t=\tfrac{3}{2}uu_{x}.\label{eq3}
\end{gather}
Equation \eqref{eq3}, often called the Riemann equation, serves as
a prototypical nonlinear partial dif\/ferential equation for the
realization of many phenomena exhibited by hyperbolic
systems~\cite{olvnut}. This might be one of the reasons why,
during the last decade, a number of works~\cite{arik} was
envisaged to study the properties of dispersionless KdV and other
related equations with special emphasis on their Lax
representation and Hamiltonian structure.

The complete integrability of the KdV equation yields the
existence of an inf\/inite family of conserved functions or
Hamiltonian densities ${{\cal H}_n}$'s that are in involution. All
${{\cal H}_n}$'s that generate f\/lows which commute with the KdV
f\/low give rise to the KdV hierarchy. The equations of the
hierarchy can be constructed using~\cite{calogero}
\begin{gather}
u_t=\Lambda^n u_x(x,t),\qquad n=0,1,2,\dots \label{eq4}
\end{gather}
with the recursion operator
\begin{gather*}
\Lambda=\tfrac{1}{4}\partial_x^2+u+\tfrac{1}{2}u_x\partial_x^{-1}.%\label{eq5}
\end{gather*}
In the dispersionless limit the recursion operator becomes
\begin{gather}
\Lambda=u+\tfrac{1}{2}u_x\partial_x^{-1}.\label{eq6}
\end{gather}
According to \eqref{eq4}, the pseudo-dif\/ferential operator
$\Lambda$ in \eqref{eq6} def\/ines a dispersionless KdV hierarchy.
The f\/irst few members of the hierarchy are given by
\begin{subequations}
\begin{gather}
n=0 :\qquad u_t=u_{x},\label{eq7a}\\
n=1 :\qquad u_t=\tfrac{3}{2}uu_{x},\label{eq7b}\\
n=2 :\qquad u_t=\tfrac{15}{8}u^2u_{x},\label{eq7c}\\
n=3 :\qquad u_t=\tfrac{35}{16}u^3u_{x},\label{eq7d}\\
%\intertext{and}
n=4 :\qquad u_t=\tfrac{315}{128}u^4u_{x}.\label{eq7e}
\end{gather}
\end{subequations}
Thus the equations in the dispersionless hierarchy can be written
in the general form
\begin{gather}
u_t=A_nu^nu_{x},\label{eq8}
\end{gather}
where the values of $A_n$ should be computed using \eqref{eq6} in
\eqref{eq4}. We can also generate $A_1$, $A_2$, $A_3$ etc
recursively using
\begin{gather*}
A_n=\left(1+\tfrac{1}{2n}\right)A_{n-1},\qquad n=1, 2, 3,\dots \qquad {\rm and}\qquad A_0=1.%\label{eq9}
\end{gather*}

The Hamiltonian structure of the dispersionless KdV hierarchy is
often studied by taking recourse to the use of Lax operators
expressed in the semi-classical limit \cite{olver}. In this work
we shall follow a dif\/ferent viewpoint to derive Hamiltonian
structure of the equations in \eqref{eq8}. We shall construct an
expression for the Lagrangian density and use the time-honoured
method of classical mechanics to rederive and reexamine the
corresponding canonical formulation. A single evolution equation
is never the Euler--Lagrange equation of a variational problem.
One common trick to put a single evolution equation into a
variational form is to replace $u$ by a potential function
$u=-w_x$. In terms of $w$, \eqref{eq8} will become an
Euler--Lagrange equation. We can, however, couple a nonlinear
evolution equation with an associated one and derive the action
principle. This allows one to write the Lagrangian density in
terms of the original f\/ield variables rather than the $w$'s,
often called the Casimir potential. In Section~2 we adapt both
these approaches to obtain the Lagrangian and Hamiltonian
densities of the Riemann type equations. In Section~3 we study the
bi-Hamiltonian structure~\cite{magri}. One of the added advantage
of the Lagrangian description is that it allows one to establish,
via Noether's theorem, the relationship between variational
symmetries and associated conservation laws. The concept of
variational symmetry results from the application of group methods
in the calculus of variations. Here one deals with the symmetry
group of an action functional ${\cal A}[u]=\int_{\Omega_0}{\cal
L}\left(x, u^{(n)}\right)dx$ with ${\cal L}$, the so-called
Lagrangian density of the f\/ield $u(x)$. The groups considered
will be local groups of transformations acting on an open subset
${\cal M}\subset \Omega_0\times U \subset X\times U$. The symbols
$X$ and $U$ denote the space of independent and dependent
variables respectively. We devote Section~4 to study this
classical problem. Finally, in Section~5 we make some concluding
remarks.

\section{Lagrangian and Hamiltonian densities}

For $u=-w_x$ \eqref{eq8} becomes
\begin{gather}
w_{xt}=A_n(-1)^nw_x^nw_{2x}.\label{eq10}
\end{gather}
The Fr\'echet derivative of the right side of \eqref{eq10} is
self-adjoint. Thus we can use the homotopy formula \cite{frankel}
to obtain the Lagrangian density in the form
\begin{gather}
{\cal L}_n=\tfrac{1}{2}w_t
w_x+\frac{A_n(-1)^{n+1}}{(n+1)(n+2)}w_x^{n+2}.\label{eq11}
\end{gather}
In writing \eqref{eq11} we have subtracted a gauge term which is
harmless at the classical level. The subscript $n$ of ${\cal L}$
merely indicates that it is the Lagrangian density for the $n$th
member of the dispersionless KdV hierarchy. The corresponding
canonical Hamiltonian densities obtained by the use of Legendre
map are given by
\begin{gather}
{\cal H}_n=\frac{A_n}{(n+1)(n+2)}u^{n+2}.\label{eq12}
\end{gather}

Equation \eqref{eq8} can be written in the form
\begin{gather}
u_t+\frac{\partial \rho[u]}{\partial x}=0\label{eq13}
\end{gather}
with
\begin{gather}
\rho[u]=-\frac{A_n}{(n+1)}u^{n+1}.\label{eq14}
\end{gather}
There exists a prolongation of \eqref{eq8} or \eqref{eq13} into
another equation
\begin{gather}
v_t+\frac{\delta(\rho[u] v_x)}{\delta u}=0,\qquad
v=v(x,t)\label{eq15}
\end{gather}
with the variational derivative
\begin{gather*}
\frac{\delta}{\delta u}=\sum_{k=0}^m(-1)^k\frac{\partial^k}{\partial x^k}\frac{\partial}{\partial u_{kx}},\qquad u_{kx}=\frac{\partial^k u}{\partial x^k}%\label{eq16}
\end{gather*}
such that the coupled system of equations follows from the action
principle \cite{bt}
\begin{gather*}
\delta\int {\cal L}^c \,dx dt=0.%\label{eq17}
\end{gather*}
The Lagrangian density for the coupled equations in \eqref{eq13}
and \eqref{eq15} is given by
\begin{gather*}
{\cal L}^c=\tfrac{1}{2}(vu_t-uv_t)-\rho[u] v_x.%\label{eq18}
\end{gather*}
For $\rho[u]$ in \eqref{eq14}, \eqref{eq15} reads
\begin{gather}
v_t=A_nu^nv_x.\label{eq19}
\end{gather}
For the system represented by \eqref{eq8} and \eqref{eq19} we have
\begin{gather}
{\cal
L}_n^c=\tfrac{1}{2}(vu_t-uv_t)+\frac{A_n}{(n+1)}u^{n+1}v_x.\label{eq20}
\end{gather}
The result in \eqref{eq19} could also be obtained using the method
of Kaup and Malomed \cite{kaup}. Referring back to the
supersymmetric KdV equation~\cite{das} we identify $v$ as a
fermionic variable associated with the bosonic equation in
\eqref{eq8}. It is of interest to note that the supersymmetric
system is complete in the sense of variational principle while
neither of the partners is. The Hamiltonian density obtained from
the Lagrangian  in \eqref{eq20} is given by
\begin{gather}
{\cal H}_n^c=-\frac{A_n}{(n+1)}u^{n+1}v_x.\label{eq21}
\end{gather}
It remains an interesting curiosity to demonstrate that the
results in \eqref{eq12} and \eqref{eq21} represent the conserved
densities of the dispersionless KdV and supersymmetric KdV
f\/lows. We demonstrate this by examinning the appropriate
bi-Hamiltonian structures of \eqref{eq8} and the pair \eqref{eq8}
and~\eqref{eq19}.

\section{Bi-Hamiltonian structure}

Zakharov and Faddeev \cite{zakfad} developed the Hamiltonian
approach to integrability of nonlinear evolution equations in one
spatial and one temporal (1+1) dimensions and Gardner
\cite{gadner}, in particular, interpreted the KdV equation as a
completely integrable Hamiltonian system with~$\partial_x$ as the
relevant Hamiltonian operator. A signif\/icant development in the
Hamiltonian theory is due to Magri \cite{magri} who realized that
integrable Hamiltonian systems have an additional structure. They
are bi-Hamiltonian, i.e., they are Hamiltonian with respect to two
dif\/ferent compatible Hamiltonian operators. A similar
consideration will also hold good for the dispersionless KdV
equations and we have
\begin{gather}
u_t=\partial_x\left(\frac{\delta {H}_{n}}{\delta
u}\right)=\tfrac{1}{2}\left(u\partial_x+\partial_xu\right)\left(\frac{\delta
{H}_{n-1}}{\delta u}\right),\qquad n=1,2,3 \dots .\label{eq22}
\end{gather}
Here
\begin{gather}
H=\int {\cal H}dx.\label{eq23}
\end{gather}
It is easy to verify that for $n=1$, \eqref{eq12}, \eqref{eq22}
and \eqref{eq23} give \eqref{eq7b}. The other equations of the
hierarchy can be obtained for $n=2,3,4,\dots$.  The operators
${\cal D}_1=\partial_x$ and ${\cal
D}_2=\tfrac{1}{2}\left(u\partial_x+\partial_xu\right)$ in
\eqref{eq22} are skew-adjoint and satisfy the Jacobi identity. The
dispersionless KdV equation, in particular, can be written in the
Hamiltonian form as
\begin{gather*}
u_t=\{u(x),H_1\}_1\qquad {\rm and}\qquad u_t=\{u(x),H_0\}_2%\label{eq24}
\end{gather*}
endowed with the Poisson structures
%\begin{subequations}
\begin{gather*}
\{u(x),u(y)\}_1={\cal D}_1\delta(x-y)%\label{eq25a}
%\end{gather}
\qquad {\rm and}\qquad
%\begin{gather}
\{u(x),u(y)\}_2={\cal D}_2\delta(x-y).%\label{eq25b}
\end{gather*}
%\end{subequations}
Thus ${\cal D}_1$ and  ${\cal D}_2$ constitute two compatible
Hamiltonian operators such that the equations obtained from
\eqref{eq8} are integrable in Liouville's sense \cite{magri}. Thus
${{\cal H}_n}$'s in \eqref{eq12} via \eqref{eq23} give the
conserved densities of \eqref{eq8}. In other words,  ${{\cal
H}_n}$'s generate f\/lows which commute with the dispersionless
KdV f\/low and give rise to an appropriate hierarchy. It will be
quite interesting to examine if a similar analysis  could also be
carried out for the supersymmetric dispersionless KdV equations.

  The pair of supersymmetric equations $u_t=u^nu_x$ and $v_t=u^nv_x$ can be written as
\begin{gather}
{\bf \eta}_t={\bf J}_1\left(\frac{\delta { H}_{n}^s}{\delta {\bf
\eta}}\right)={\bf J}_2\left(\frac{\delta { H}_{n-1}^s}{\delta
{\bf \eta}}\right),\label{eq26}
\end{gather}
where ${\bf
\eta}=\left(\!\begin{array}{c}u\\v\end{array}\!\right)$,
$H^s_n={H^c_n\over A_n}$ and $H^c_n=\int{\cal H}_{n}^cdx$. In
\eqref{eq26} ${\bf J}_1$ and ${\bf J}_2$  stand for the  matrices
\begin{gather}
{\bf J}_1=\left(\begin{array}{cc}
0&1\\
-1&0
\end{array}\right)\qquad {\rm and}\qquad {\bf J}_2=\left(\begin{array}{cc}
0&u \\-u&0\end{array}\right).\label{eq27}
\end{gather}
Since $H_n^c$ for dif\/ferent values of $n$ represent the
conserved Hamiltonian densities obtained by the use of action
principle, the supersymmetric dispersionless KdV equations will be
bi-Hamiltonian provided ${\bf J}_1$ and ${\bf J}_2$ constitute a
pair of compatible Hamiltonian operators. Clearly, ${\bf J}_1$ and
${\bf J}_2$ are skew-adjoint. Thus ${\bf J}_1$ and ${\bf J}_2$
will be Hamiltonian operators provided we can show that
\cite{olver}
\begin{gather}
{\rm pr\, v}_{{\bf J}_i\theta}(\Theta_{{\bf J}_i})=0,\qquad
i=1,2.\label{eq28}
\end{gather}
Here pr stands for the prolongation of the evolutionary vector
f\/ield v of the characteristic ${\bf J}_i\theta$. The quantity
${\rm pr \,v}_{{\bf J}_i}\theta$ is calculated by using
\begin{gather}
{\rm pr \,v}_{{\bf J}_i\theta}=\sum_{\mu,
j}D_j\left(\sum_{\nu}({\bf
J}_i)_{\mu\nu}\theta^\nu\right)\frac{\partial}{\partial\eta^\mu_j},\qquad
D_j=\frac{\partial}{\partial x^j}, \qquad \mu,\nu=1,2.\label{eq29}
\end{gather}
In our case the column matrix $\theta=\left(\!\begin{array}{c}
\phi\\
\psi
\end{array}\!\right)$ represents the basis univectors associated with the variables $\eta=\left(\!\begin{array}{c}
u\\
v
\end{array}\!\right)$. Understandably, $\theta^\nu$ and $\eta^\mu$ denote the components of $\theta$ and $\eta$ and $({\bf J}_i)_{\mu\nu}$ carries a similar meaning. The functional bivectors corresponding to the operators ${\bf J}_i$ is given~by
\begin{gather}
{\Theta_{{\bf J}_i}}=\tfrac{1}{2}\int\theta^T\wedge {\bf
J}_i\theta dx\label{eq30}
\end{gather}
with $\theta^T$, the transpose of $\theta$. From \eqref{eq27},
\eqref{eq29} and \eqref{eq30} we found that both ${\bf J}_1$ and
${\bf J}_2$ satisfy~\eqref{eq28} such that each of them
constitutes a Hamiltonian operator. Further, one can check that
${\bf J}_1$ and~${\bf J}_2$ satisfy the compatibility condition
\begin{gather*}
{\rm pr \,v}_{{\bf J}_1\theta}(\Theta_{{\bf J}_2})+{\rm pr \,v}_{{\bf J}_2\theta}(\Theta_{{\bf J}_1})=0.%\label{eq31}
\end{gather*}
This shows that \eqref{eq26} gives the bi-Hamiltonian form of
supersymmetric dispersionless KdV equations. The recursion
operator def\/ined by
\begin{gather*}
\Lambda={\bf J}_2{\bf J}_1^{-1}=\left(\begin{array}{cc}
u &0\\0&u\end{array}\right)%\label{eq32}
\end{gather*}
reproduces the hierarchy of supersymmetric dispersionless KdV
equation according to
\begin{gather*}
{\bf \eta}_t={\Lambda}^n{\bf \eta}_x.%\label{eq33}
\end{gather*}
for $n=0,1,2,\dots$. This verif\/ies that ${{\cal H}_{n}^c\over
A_n}$'s as conserved densities generate f\/lows which commute with
the supersymmetric dispersionless KdV f\/low.

\section{Variational symmetries}

The Lagrangian and Hamiltonian formulations of dynamical systems
give a way to make the relation between symmetries and conserved
quantities more precise and thereby provide a method to derive
expressions for the conserved quantities from the symmetry
transformations. In its general form this is referred to as
Noether's theorem. More precisely, this theorem asserts that if a
given system of dif\/ferential equations follows from the
variational principle, then a continuous symmetry transformation
(point, contact or higher order) that leaves the action functional
invariant to within a divergence yields a conservation law. The
proof of this theorem requires some knowledge of dif\/ferential
forms, Lie derivatives and pull-back~\cite{olver}. We shall,
however, carry out the symmetry analysis for the dispersionless
KdV equation using a relatively simpler mathematical framework as
compared to that of the algebro-geometric theories. In fact, we
shall make use of some point transformations that depend on time
and spatial coordinates. The approach to be followed by us has an
old root in the classical-mechanics literature. For example, as
early as 1951, Hill \cite{hill} provided a simplif\/ied account of
Noether's theorem by considering inf\/initesimal transformations
of the dependent and independent variables characterizing the
classical f\/ield. We shall f\/irst present our general scheme for
symmetry analysis and then study the variational or Noether's
symmetries of the dispersionless KdV equation.

 Consider the inf\/initesimal transformations
 \begin{subequations}\label{eq34}
\begin{gather}
{x^i}'=x^i+ \delta x^i,\qquad \delta
x^i=\epsilon\xi^i(x,f)\label{eq34a}
\end{gather}
and
\begin{gather}
{f}'=f+ \delta f,\qquad \delta f=\epsilon\eta(x,f)\label{eq34b}
\end{gather}
\end{subequations}
for a f\/ield variable $f=f(x,t)$ with $\epsilon$, an arbitrary
small quantity. Here $x=\{x^0,x^1\}$, $x^0=t$ and $x^1=x$.
Understandably, our treatment for the symmetry analysis will be
applicable to $(1+1)$ dimensional cases. However, the result to be
presented here can easily be generalized to deal with $(3+1)$
dimensional problems. For an arbitrary analytic function
$g=g(x^i,f)$, it is straightforward to show that
\begin{gather*}
\delta g=\epsilon X g%\label{eq35}
\end{gather*}
with
\begin{gather}
X =\xi^i\frac{\partial}{\partial x^i}+\eta\frac{\partial}{\partial
f},\label{eq36}
\end{gather}
the generator of the inf\/initesimal transformations in
\eqref{eq34}. A similar consideration when applied to
$h=h(x^i,f,f_i)$ with $f_i=\frac{\partial f}{\partial x^i}$ gives
\begin{gather}
\delta h=\epsilon X' h\label{eq37}
\end{gather}
with
\begin{gather}
X'=X+\big(\eta_i-\xi^j_if_j\big)\frac{\partial}{\partial
f_i}.\label{eq38}
\end{gather}
Understandably, $X'$ stands for the f\/irst prolongation of $X$.
To arrive at the statement for the Noether's theorem we consider
among the general set of transformations in \eqref{eq34} only
those that leave the f\/ield-theoretic action invariant. We thus
write
\begin{gather}
{\cal L}(x^i,f,f_i)d(x)={\cal
L'}({x^i}',{f}',{f_i}')d(x'),\label{eq39}
\end{gather}
where $d(x)=dxdt$. In order to satisfy the condition in
\eqref{eq39} we allow the Lagrangian density to change its
functional form ${\cal L}$ to ${\cal L'}$. If the equations of
motion, expressed in terms of the new variables, are to be of
precisely the same functional form as in the old variables, the
two density functions must be related by a divergence
transformation. We  thus express the relation between~${\cal L'}$
and~${\cal L}$ by introducing a gauge function $B^i(x,f)$ such
that
\begin{gather}
{\cal L'}({x^i}',{f}',{f_i}')d(x')={\cal
L}({x^i}',{f}',{f_i}')d(x')-\epsilon\frac{dB^i}{d{x^i}'}d(x')+o(\epsilon^2).\label{eq40}
\end{gather}
The general form of \eqref{eq40} for the def\/inition of symmetry
transformations will allow the scale and divergence
transformations to be considered as symmetry transformations.
Understandably, the scale transformations give rise to Noether's
symmetries while the scale transformations in conjunction with the
divergence term lead to Noether's divergence symmetries.
Traditionally, the concept of divergence symmetries and
concommitant conservation laws are introduced by replacing
Noether's inf\/initesimal criterion for invariance by a divergence
condition \cite{gelfand}. However, one can directly work with the
conserved densities that follow from \eqref{eq40} because nature
of the vector f\/ields will determine the contributions of the
gauge term. For some of the vector f\/ields the contributions of
$B^i$ to conserved quantities will be equal to zero. These vector
f\/ields are Noether's symmetries else we have Noether's
divergence symmetries. In view of \eqref{eq39}, \eqref{eq40} can
be written in the form
\begin{gather}
{\cal L}({x^i}',{f}',{f_i}')d(x')={\cal
L}({x^i},{f},{f_i})d(x)+\epsilon\frac{dB^i}{d{x^i}}d(x).\label{eq41}
\end{gather}
Again using ${\cal L}$ for $h$ in \eqref{eq37}, we have
\begin{gather}
{\cal L}({x^i}',{f}',{f_i}')d(x')={\cal
L}({x^i},{f},{f_i})\left[d(x)+\epsilon
d\xi^i(x,f_i)\right]+\epsilon X'{\cal
L}({x^i},{f},{f_i})d(x).\label{eq42}
\end{gather}
From \eqref{eq41} and \eqref{eq42}, we write
\begin{gather}
\frac{dB^i}{d{x^i}}=\frac{d\xi^i}{d{x^i}}{\cal L}+X'{\cal
L}.\label{eq43}
\end{gather}
Using the value of $X'$ from \eqref{eq38} in \eqref{eq43},
$\frac{dB^i}{d{x^i}}$ is obtained in the f\/inal form
\begin{gather}
\frac{dB^i}{d{x^i}}=\frac{d\xi^i}{d{x^i}}{\cal
L}+\xi^i\frac{\partial {\cal L}}{\partial x^i}+\eta\frac{\partial
{\cal L}}{\partial f}+\big(\eta_i-\xi^j_if_j\big)\frac{\partial
{\cal L}}{\partial f_i}.\label{eq44}
\end{gather}
Thus we f\/ind that the action is invariant under those
transformations whose constituents~$\xi$ and~$\eta$ satisfy
\eqref{eq44}. The terms in \eqref{eq44} can be rearranged to write
\begin{gather}
\frac{d}{d{x^i}}\left\{B^i-\xi^i{\cal
L}+\left(\xi^jf_j-\eta\right)\frac{\partial {\cal L}}{\partial
f_i}\right\}+\left(\xi^jf_j-\eta\right)\left[\frac{\partial {\cal
L}}{\partial f}-\frac{d}{d{x^i}}\left(\frac{\partial {\cal
L}}{\partial f_i}\right)\right]=0.\label{eq45}
\end{gather}
The expression inside the squared bracket stands for the
Euler--Lagrange equation for the classical f\/ield under
consideration. In view of this, \eqref{eq45} leads to the
conservation law
\begin{gather}
\frac{d{\cal I}^i}{d{x^i}}=0\label{eq46}
\end{gather}
with the conserved density given by
\begin{gather}
{\cal I}^i=B^i-\xi^i{\cal
L}+\left(\xi^jf_j-\eta\right)\frac{\partial {\cal L}}{\partial
f_i}.\label{eq47}
\end{gather}
In the case of two independent variables $(x^0,x^1)\equiv(t,x)$,
\eqref{eq46} can be written in the explicit form
\begin{gather}
\frac{d{\cal I}^0}{d{t}}+\frac{d{\cal I}^1}{d{x}}=0.\label{eq48}
\end{gather}

From \eqref{eq11} the Lagrangian density for the dispersionless
KdV equation is obtained as
\begin{gather}
{\cal L}=\tfrac{1}{2}w_tw_x+\tfrac{1}{4}w_x^3.\label{eq49}
\end{gather}
Identifying $f$ with $w$ we can combine \eqref{eq47}, \eqref{eq48}
and \eqref{eq49} to get
\begin{gather}
B^0_t+w_tB^0_w-\tfrac{1}{4}\xi^0_tw_x^3-\tfrac{1}{4}\xi^0_ww_tw_x^3+\tfrac{1}{2}\xi_t^1w_x^2
+\tfrac{1}{2}\xi^1_ww_tw_x^2-\tfrac{1}{2}\eta_tw_x-\eta_ww_tw_x\nonumber\\
\qquad{}{}+B^1_x+w_xB^1_w+\tfrac{1}{2}\xi^1_xw_x^3+\tfrac{1}{2}\xi^1_ww_x^4+\tfrac{1}{2}\xi^0_xw_t^2+
\tfrac{1}{2}w_xw_t^2\xi^0_w+\tfrac{3}{4}\xi^0_xw_x^2w_t-\tfrac{3}{4}\eta_xw_x^2\nonumber\\
\qquad{}{}-\tfrac{1}{2}\eta_xw_t-
\tfrac{3}{4}\eta_ww_x^3+\tfrac{3}{4}\xi_w^0w_x^3w_t=0.\label{eq50}
\end{gather}
In writing \eqref{eq50} we have made use of \eqref{eq10} with
$n=1$. Equation \eqref{eq50} can be globally satisf\/ied if\/f the
coef\/f\/icients of the following terms vanish separately
\begin{subequations}\label{eq51}
\begin{gather}
w_{x}^0\quad\mbox{or}\quad w_{t}^0:\qquad B^0_t+B^1_x=0,\label{eq51a}\\
w_t:\qquad B^0_w-\tfrac{1}{2}\eta_x=0,\label{eq51b}\\
w_t^2:\qquad \tfrac{1}{2}\xi^0_x=0,\label{eq51c}\\
w_x:\qquad B^1_w-\tfrac{1}{2}\eta_t=0,\label{eq51d}\\
w_x^2:\qquad \tfrac{1}{2}\xi_t^1-\tfrac{3}{4}\eta_x=0,\label{eq51e}\\
w_x^3:\qquad -\tfrac{1}{4}\xi^0_t-\tfrac{3}{4}\eta_w+\tfrac{1}{2}\xi^1_x=0,\label{eq51f}\\
w_x^4:\qquad \tfrac{1}{2}\xi^1_w=0,\label{eq51g}\\
w_tw_x:\qquad -\eta_w=0,\label{eq51h}\\
w_tw_x^2:\qquad \tfrac{1}{2}\xi^1_w+\tfrac{3}{4}\xi^0_x=0,\label{eq51i}\\
w_tw_x^3:\qquad \tfrac{1}{2}\xi^0_w=0,\label{eq51j}\\
%and
w_t^2w_x:\qquad \tfrac{1}{2}\xi^0_w=0.\label{eq51k}
\end{gather}
\end{subequations}
Equations in \eqref{eq51} will lead to f\/inite number of
symmetries. This number appears to be disappointingly small since
we have a dispersionless KdV hierarchy given in \eqref{eq8}.
Further, symmetry properties ref\/lecting the existence of
inf\/initely many conservation laws will require an appropriate
development for the theory of generalized symmetries. In this
work, however, we shall be concerned with variational symmetries
only.

From \eqref{eq51c}, \eqref{eq51j} and \eqref{eq51k} we see that
$\xi^0$ is only a function of $t$. We, therefore, write
\begin{gather}
\xi^0(x,t,w)=\beta(t).\label{eq52a}
\end{gather}
Also from \eqref{eq51g}, \eqref{eq51i} and \eqref{eq52a} we see
that $\xi^1$ is not a function of $w$. In view of \eqref{eq51h}
and \eqref{eq52a}, \eqref{eq51f} gives
\begin{gather*}
\xi^1_x-\tfrac{1}{2}\beta_t=0%\label{eq53a}
\end{gather*}
which can be solved to get
\begin{gather}
\xi^1=\tfrac{1}{2}\beta_tx+\alpha(t),\label{eq52b}
\end{gather}
where $\alpha(t)$ is a constant of integration. Using
\eqref{eq52b} in \eqref{eq51e} we have
\begin{gather}
\eta_x=\tfrac{1}{3}\beta_{tt}x+\tfrac{2}{3}\alpha_t.\label{eq53b}
\end{gather}
The solution of \eqref{eq53b} is given by
\begin{gather}
\eta=\tfrac{1}{6}\beta_{tt}x^2+\tfrac{2}{3}\alpha_tx+\gamma(t)\label{eq52c}
\end{gather}
with $\gamma(t)$, a constant of integration. In view of
\eqref{eq52c}, \eqref{eq51b} and \eqref{eq51d} yield
\begin{gather}
B^0=\tfrac{1}{6}\beta_{tt}xw+\tfrac{1}{3}\alpha_tw\label{eq54}
\end{gather}
and
\begin{gather}
B^1=\tfrac{1}{12}\beta_{ttt}x^2w+\tfrac{1}{3}\alpha_{tt}xw.\label{eq55}
\end{gather}
Equations \eqref{eq54} and \eqref{eq55} can be combined with
\eqref{eq51a} to get f\/inally
\begin{gather}
\beta_{ttt}=0\qquad {\rm and}\qquad \alpha_{tt}=0.\label{eq56}
\end{gather}
From \eqref{eq56} we write
\begin{gather}
\beta=\tfrac{1}{2}a_1t^2+a_2t+a_3\label{eq57}
\end{gather}
and
\begin{gather}
\alpha=b_1t+b_2,\label{eq58}
\end{gather}
where $a$'s and $b$'s are arbitrary constants. Substituting the
values of $\beta$ and $\alpha$ in \eqref{eq52a},
  \eqref{eq52b}, \eqref{eq52c} we obtain the inf\/initesimal transformation, $\xi^0$, $\xi^1$ and $\eta$, as
\begin{subequations}\label{eq59}
\begin{gather}
\xi^0=\tfrac{1}{2} a_1 t^2+a_2t+a_3,\label{eq59a}\\
\xi^1=\tfrac{1}{2}(a_1t+a_2)x+b_1t+b_2,\label{eq59b}\\
%and
\eta=\tfrac{1}{6}a_1x^2+\tfrac{2}{3}b_1x+b_3.\label{eq59c}
\end{gather}
\end{subequations}
In writing \eqref{eq59c} we have treated $\gamma(t)$ as a constant
and replaced it by $b_3$. Implication of this choice will be made
clear while considering the symmetry algebra. In terms of
\eqref{eq59}, \eqref{eq36} becomes
\begin{gather*}
X=a_1V_1+a_2V_2+a_3V_3+b_1V_4+b_2V_5+b_3V_6,%\label{eq60}
\end{gather*}
where
%\begin{subequations}\label{eq61}
\begin{gather}
V_1=\tfrac{1}{2}t^2\frac{\partial}{\partial
t}+\tfrac{1}{2}xt\frac{\partial}{\partial
x}+\tfrac{1}{6}x^2\frac{\partial}{\partial w},
\qquad %\label{eq61a}\\
V_2=t\frac{\partial}{\partial t}+\tfrac{1}{2}x\frac{\partial}{\partial x},\nonumber\\ %\label{eq61b}\\
V_3=\frac{\partial}{\partial t},\qquad %\label{eq61c}\\
V_4=t\frac{\partial}{\partial x}+\tfrac{2}{3}x\frac{\partial}{\partial w}, \qquad %\label{61d}\\
V_5=\frac{\partial}{\partial x},\qquad  %\label{eq61e}\\
%and
V_6=\frac{\partial}{\partial w}.\label{eq61}%\label{eq61f}
\end{gather}
%\end{subequations}
It is easy to check that the vector f\/ields $V_1,\dots,V_6$
satisfy the closure property. The commutation relations between
these vector f\/ields are given in Table~\ref{table1}.

\begin{table}[h]
\centering

\caption{Commutation relations for the generators in \eqref{eq61}.
Each element $V_{ij}$ in the Table is represented by
$V_{ij}=[V_i,\,V_j]$.}

\begin{tabular}{c|cccccc}
&$V_1$&$V_2$ &$V_3$&$V_4$&$V_5$&$V_6$ \\
\hline
$V_1$&$0$&$-V_1$ &$-V_2$&$0$&$-{1\over 2}V_4$&$0$\\
$V_2$&$V_1$&$0$ &$-V_3$&${1\over 2}V_4$&$-{1\over 2}V_5$&$0$\\
$V_3$&$V_2$&$V_3$&$0$&$V_5$&$0$&$0$\\
$V_4$&$0$&$-{1\over 2}V_4$&$-V_5$&$0$&$-{2\over 3}V_6$&$0$\\
$V_5$&${1\over 2}V_4$&${1\over 2}V_5$ &$0$&${2\over 3}V_6$&$0$&$0$\\
$V_6$&$0$&$0$&$0$&$0$&$0$&$0$
\end{tabular}\label{table1}
\end{table}

The symmetries in \eqref{eq61} are expressed in terms of the
velocity f\/ield and depend explicitly on $x$ and $t$. Looking
from this point of view the symmetry vectors obtained by us bear
some similarity with the so called `addition symmetries' suggested
independently by Chen, Lee and Lin \cite{chenlee} and by Orlov and
Shulman \cite{orlov}. It is easy to see that $V_2$ to $V_6$
correspond to scaling, time translation, Galilean boost, space
translation and translation in velocity space respectively. The
vector f\/ield $V_1$ does not admit such a simple physical
realization. However, we can write~$V_1$ as
$V_1={1\over2}tV_2+{1\over4}xV_4$.

Making use of \eqref{eq49}, \eqref{eq54}, \eqref{eq55},
\eqref{eq57} and \eqref{eq58} we can write the expressions for the
conserved quantities in \eqref{eq47} as
\begin{subequations}
\begin{gather}
{\cal I}^0=\tfrac{1}{6}a_1xw+\tfrac{1}{3}b_1w-\tfrac{1}{4}\xi^0w_x^3+\tfrac{1}{2}\xi^1w_x^2-\tfrac{1}{2}\eta w_x,\label{eq62a}\\
%and
{\cal
I}^1=\tfrac{1}{2}\xi^0w_t^2+\tfrac{3}{4}\xi^0w_tw_x^2+\tfrac{1}{2}\xi^1w_x^3-\tfrac{1}{2}\eta
w_t-\tfrac{3}{4}\eta w_x^2.\label{eq62b}
\end{gather}
\end{subequations}
The expressions for ${\cal I}^0$ and ${\cal I}^1$ are
characterized by $\xi^i$ and $\eta$, the values of which change as
we go from one vector f\/ield to the other. The f\/irst two terms
in ${\cal I}^0$ stand for the contribution of~$B^0$ and there is
no contribution of the gauge term in ${\cal I}^1$ since from
\eqref{eq55} and \eqref{eq56} $B^1=0$. For a~particular vector
f\/ield $a_1$ and $b_1$ may either be zero or non zero. One can
verify that except for vector f\/ields $V_1$ and $V_4$,
$a_1=b_1=0$ such that $V_2$, $V_3$, $V_5$ and $V_6$ are simple
Noether's symmetries while $V_1$ and $V_4$ are Noether's
divergence symmetries. Coming down to details we have found the
following conserved quantities from \eqref{eq62a} and
\eqref{eq62b}
\begin{subequations}\label{eq63}
\begin{gather}
{\cal I}^0_{V_1}=\tfrac{1}{6}xw-\tfrac{1}{8}t^2w_x^3+\tfrac{1}{4}xtw_x^2-\tfrac{1}{12}x^2w_x,\label{eq63a}\\
{\cal I}^1_{V_1}= \tfrac{1}{4}xtw_x^3+ \tfrac{3}{8}t^2w_tw_x^2+ \tfrac{1}{4}t^2w_t^2- \tfrac{1}{12}x^2w_t- \tfrac{1}{8}x^2w_x^2,\label{eq63b}\\
{\cal I}^0_{V_2}=-\tfrac{1}{4}tw_x^3+\tfrac{1}{4}xw_x^2,\label{eq63c}\\
{\cal I}^1_{V_2}=\tfrac{1}{4}xw_x^3+\tfrac{3}{4}tw_tw_x^2+\tfrac{1}{2}tw_t^2,\label{eq63d}\\
{\cal I}^0_{V_3}=-\tfrac{1}{4}w_x^3,\label{eq63e}\\
{\cal I}^1_{V_3}=\tfrac{3}{4}w_tw_x^2+\tfrac{1}{2}w_t^2,\label{eq63f}\\
{\cal I}^0_{V_4}=\tfrac{1}{3}w+\tfrac{1}{2}tw_x^2-\tfrac{1}{3}xw_x,\label{eq63g}\\
{\cal I}^1_{V_4}=\tfrac{1}{2}tw_x^3-\tfrac{1}{3}xw_t-\tfrac{1}{2}xw_x^2,\label{eq63h}\\
{\cal I}^0_{V_5}=\tfrac{1}{2}w_x^2,\label{eq63i}\\
{\cal I}^1_{V_5}=\tfrac{1}{2}w_x^3,\label{eq63j}\\
{\cal I}^0_{V_6}=-\tfrac{1}{2}w_x,\label{eq63k}\\
%and
{\cal I}^1_{V_6}=-\tfrac{1}{2}w_t-\tfrac{3}{4}w_x^2.\label{eq63l}
\end{gather}
\end{subequations}
It is easy to check that the results in \eqref{eq63}  is
consistent with \eqref{eq48}. The pair of conserved quantities
corresponding to time translation, space translation and velocity
space translation, namely, \{\eqref{eq63e},\eqref{eq63f}\},
\{\eqref{eq63i},\eqref{eq63j}\} and
\{\eqref{eq63k},\eqref{eq63l}\} do not involve $x$ and $t$
explicitly. Each of the pair in conjunction with \eqref{eq48} give the
dispersionless KdV equation in a rather straightforward manner. As
expected \eqref{eq63e} stands for the Hamiltonian density or
energy of~\eqref{eq7b}.

\section{Conclusion}
Compatible Hamiltonian structures of the dispersionless KdV
hierarchy are traditionally obtained with special attention to
their Lax representation in the semiclassical limit. The
derivation involves judicious use of the so-called $r$-matrix
method~\cite{takha}. We have shown that the combined Lax
representation--$r$-matrix method can be supplemented by a Lagrangian
approach to the problem. We found that the Hamiltonian densities
corresponding to our Lagrangian representations stand for the
conserved densities for the dispersionless KdV f\/low. We could
easily construct the Hamiltonian operators from the recursion
operator which generates the hierarchy. We have derived the
bi-Hamiltonian structures for both dispersionless KdV and
supersymmetric KdV hierarchies. As an added realism of the
Lagrangian approach we studied the variational symmetries of
equation \eqref{eq7b}. We believe that it will be quite
interesting to carry out similar analysis for the supersymmetric
KdV pair in \eqref{eq7b} and for $n=1$ limit of~\eqref{eq19}.

\subsection*{Acknowledgements}

This work is supported by the University Grants Commission,
Government of India, through grant No. F.32-39/2006(SR).

\pdfbookmark[1]{References}{ref}

\LastPageEnding


\begin{thebibliography}{99}

\footnotesize\itemsep=0pt

\bibitem{zakharov} Zakharov V.E., Benney equations and quasiclassical approximation in the method of the inverse problem, {\it Funct. Anal. Appl.} {\bf 14} (1980), 89--98.

\bibitem{olvnut} Olver P.J., Nutku Y., Hamiltonian structures for systems of hyperbolic conservation laws, {\it J. Math. Phys.} {\bf 29} (1988),  1610--1619.\\
     Brunelli J.C., Dispersionless limit of integrable models, {\it Braz. J. Phys.} {\bf 30} (2000), 455--468, \href{http://arxiv.org/abs/nlin.SI/0207042}{nlin.SI/0207042}.

\bibitem{arik} Arik M., Neyzi F.,  Nutku Y., Olver P.J., Verosky J.M., Multi-Hamiltonian structure of the Born--Infeld equation, {\it J. Math. Phys.} {\bf 30} (1989), 1338--1344.\\
      Das A., Huang W.J., The Hamiltonian structures associated with a generalized Lax operator, {\it J. Math. Phys.} {\bf 33} (1992), 2487--2497.\\
        Brunelli J.C., Das A., Properties of nonlocal charges in the supersymmetric two boson hierarchy, {\it Phys. Lett.~B} {\bf 354} (1995), 307--314, \href{http://arxiv.org/abs/hep-th/9504030}{hep-th/9504030}.\\
        Brunelli J.C.,  Das A., Supersymmetric two-boson equation, its reductions and the nonstandard supersymmetric KP hierarchy,  {\it Intern. J. Modern Phys.~A} {\bf 10} (1995), 4563--4599, \href{http://arxiv.org/abs/hep-th/9505093}{hep-th/9505093}.\\
Brunelli J.C., Hamiltonian structures for the generalized dispersionless KdV hierarchy, {\it Rev. Math. Phys.} {\bf 8} (1996), 1041--1054, \href{http://arxiv.org/abs/solv-int/9601001}{solv-int/9601001}.\\
Brunelli J.C., Das A., A Lax description for polytropic gas dynamics, {\it Phys. Lett.~A} {\bf 235} (1997), 597--602, \href{http://arxiv.org/abs/solv-int/9706005}{solv-int/9706005}.\\
Brunelli J.C., Das A., The sTB-B hierarchy, {\it Phys. Lett. B} {\bf 409} (1997),  229--238, \href{http://arxiv.org/abs/hep-th/9704126}{hep-th/9704126}.\\
Brunelli J.C., Das A.,  A Lax representation for Born--Infeld
equation, {\it Phys. Lett. B} {\bf 426} (1998), 57--63,
\href{http://arxiv.org/abs/hep-th/9712081}{hep-th/9712081}.

 \bibitem{calogero} Calogero F., Degasperis A., Spectral transform and soliton, North-Holland Publising Company, New York, 1982.

\bibitem{olver} Olver  P.J., Application of Lie groups to dif\/ferential equation, Springer-Verlag, New York, 1993.

\bibitem{magri}  Magri F., A simple model of the integrable Hamiltonian equation,  {\it J. Math. Phys.} {\bf 19} (1978),  1156--1162.

\bibitem{frankel} Frankel T., The geometry of physics, Cambridge University Press, UK, 1997.

\bibitem{bt} Ali Sk.G., Talukdar  B., Das U., Inverse problem of variational calculus for nonlinear evolution equations, {\it Acta Phys. Polon. B} {\bf 38} (2007), 1993--2002, \href{http://arxiv.org/abs/nlin.SI/0603037}{nlin.SI/0603037}.

\bibitem{kaup} Kaup D.J., Malomed B.A., The variational principle for nonlinear waves in dissipative systems, {\it Phys.~D} {\bf 87} (1995), 155--159.

\bibitem{das} Barcelos-Neto J., Constandache A., Das A., Dispersionless fermionic KdV, {\it Phys. Lett.~A} {\bf 268} (2000), 342--351,
\href{http://arxiv.org/abs/solv-int/9910001}{solv-int/9910001}.

\bibitem{zakfad} Zakharov V.E., Faddeev L.D., Korteweg--de Vries equation: a completely integrable Hamiltonian systems, {\it Funct. Anal. Appl.} {\bf 5} (1971),  18--27.

\bibitem{gadner} Gardner C.S., Korteweg--de Vries equation and generalizations. IV. The Korteweg--de Vries equation as a~Hamiltonian system, {\it J. Math. Phys.} {\bf 12} (1971), 1548--1551.

\bibitem{hill} Hill E.L., Hamilton's principle and the conservation theorems of mathematical physics, {\it Rev. Modern Phys.} {\bf 23} (1951),  253--260.

\bibitem{gelfand} Gelfand I.M., Fomin S.V., Calculus of variations, Dover Publ., 2000.

\bibitem{chenlee} Chen H.H., Lee Y.C., Lin J.E., On a new hierarchy of symmetries for the Kadomtsev--Petviashvilli equation, {\it Phys.~D} {\bf 9} (1983), 439--445.

\bibitem{orlov} Orlov A.Yu., Shulman E.I., Additional symmetries for integral and conformal algebra representation, {\it Lett. Math. Phys.} {\bf 12}  (1986), 171--179.

\bibitem{takha} Faddeev  L.D., Takhtajan L.A., Hamiltonian methods in the theory of solitons, Springer, Berlin, 1987.
\end{thebibliography}
\end{document}